\documentclass{nature}
\usepackage{graphicx}
\makeatletter
\let\saved@includegraphics\includegraphics
\AtBeginDocument{\let\includegraphics\saved@includegraphics}
\renewenvironment*{figure}{\@float{figure}}{\end@float}
\makeatother

\usepackage{multirow}
\usepackage{subfigure}
\usepackage{mathtools}
\usepackage[colorlinks=true,linkcolor=blue,citecolor=black,urlcolor=blue]{hyperref}
\usepackage{authblk}
\usepackage[labelfont=bf]{caption}
\usepackage[figurename=Figure]{caption}
\usepackage{siunitx}
\usepackage[dvipsnames]{xcolor}
\usepackage{soul}
\usepackage{xfrac}
\usepackage{amsmath}
\usepackage{soul,xcolor}
\usepackage{lineno}
\usepackage{siunitx}
\graphicspath{{./figures/}}

\setstcolor{red}

\title{Current-driven dynamics of antiferromagnetic skyrmions: from skyrmion Hall effects to hybrid inter-skyrmion scattering} 
\author{Amal Aldarawsheh$^{1,2*}$, Moritz Sallermann$^{1,3,4}$,Muayad Abusaa$^5$ and  Samir Lounis$^{1,2*}$}
\begin{document}
	\maketitle
	\begin{affiliations}
		\item Peter Gr\"{u}nberg Institute and Institute for Advanced Simulation, Forschungszentrum J\"{u}lich and JARA, D-52425 J\"{u}lich, Germany
		\item Faculty of Physics, University of Duisburg-Essen and CENIDE, 47053 Duisburg, Germany
		\item RWTH Aachen University, 52056 Aachen, Germany
		\item Science Institute and Faculty of Physical Sciences, University of Iceland, VR-III, 107 Reykjavík, Iceland
		\item Department of Physics, Arab American University, Jenin, Palestine\
		
		* a.aldarawsheh@fz-juelich.de; s.lounis@fz-juelich.de
		
	\end{affiliations}
	\section*{Abstract}
Antiferromagnetic (AFM) skyrmions 
 have emerged as a highly promising avenue in the realm of spintronics, particularly for the development of advanced racetrack memory devices. A distinguishing feature of AFM skyrmions is their zero topological charge and hence  anticipated zero skyrmion Hall effect (SkHE). Here, we unveil that the latter is surprisingly finite under the influence of spin-transfer torque, depending on the direction of the injected current impinging on intrinsic AFM skyrmions emerging in CrPdFe trilayer on Ir(111) surface. Hinging on first-principles combined with atomistic spin dynamics simulations, we identify the origin of the SkHE and  uncover that FM skyrmions in the underlying Fe layer act as effective traps for AFM skyrmions, confining them and reducing their velocity. These findings hold significant promise for spintronic applications, the design of multi-purpose skyrmion-tracks, advancing our understanding of AFM-FM skyrmion interactions and hybrid soliton dynamics in heterostructures.

\section{Introduction}

Topologically protected magnetic textures, such as magnetic skyrmions~\cite{roessler2006spontaneous, fert2013skyrmions,Nagaosa2013,finocchio2016magnetic,Fert2017,everschor2018perspective,zhou2019magnetic,zhang2020skyrmion}, have emerged as a subject of immense interest due to their potential to become non-volatile information carriers in future spintronic devices~\cite{zhang2015magnetic,kang2016skyrmion,prychynenko2018magnetic,nozaki2019brownian}. The discovery of magnetic skyrmions in numerous materials hosting bulk or interfacial Dzyaloshinskii-Moriya interaction (DMI) has spurred extensive research to explore their unique properties and applications~\cite{Nagaosa2013,finocchio2016magnetic,Fert2017,everschor2018perspective}. Particularly, the concept of skyrmion-based racetrack memory~\cite{fert2013skyrmions,Nagaosa2013,iwasaki2013current,zhang2015skyrmion} has garnered considerable attention, where skyrmions are dynamically driven by numerous means. These skyrmions possess several  advantages, including nanoscale size, topological stability,  ability to overcome pinning sites~\cite{romming2013writing,Woo2016,legrand2017room,fernandes2018universality,fernandes2020impurity,Arjana2020}, and low depinning current compared to domain walls~\cite{jonietz2010spin,miron2011fast,yu2012skyrmion,yang2015domain}.%, making them promising candidates for next-generation spintronics.

Various methods have been proposed to drive magnetic skyrmions, encompassing electric currents~\cite{fert2013skyrmions,iwasaki2013current,zhang2015skyrmion,Sampaio2013}, spin waves~\cite{zhang2015all}, magnetic field gradients ~\cite{komineas2015skyrmion}, temperature gradients~\cite{kong2013dynamics}, and voltage-controlled magnetic anisotropy~\cite{ma2018electric,tomasello2018chiral,xia2018skyrmion,wang2018efficient}. However, one significant challenge that arises during their manipulation via electrical means or magnetic field gradient~\cite{liang2018magnetic} is the skyrmion Hall effect (SkHE), wherein skyrmion trajectories deviate from the driving current direction due to the Magnus force~\cite{cortes2017thermal,everschor2014real,jiang2017direct} proportional to the topological charge~\cite{Nagaosa2013}.  This undesired effect hampers the precise control and movement of skyrmions in spintronic devices.

In contrast,  antiferromagnetic (AFM) skyrmions are expected to be transparent to the SkHE since the building-blocks skyrmions carry opposite topological charge, which enforce the motion along the  direction imposed by the applied current, as predicted  theoretically ~\cite{zhang2016antiferromagnetic,zhang2016magnetic,barker2016static} and observed for significant distances experimentally~\cite{dohi2019formation}. Moreover, AFM skyrmions  are in general insensitive to external magnetic fields~\cite{rosales2015three,dos2020modeling,bessarab2019stability,potkina2020skyrmions,aldarawsheh2022emergence,Aldarawsheh2023}, which overall promotes their discovery and control for promising implementation in devices.
Following extensive phenomenology-based predictions~\cite{zhang2016antiferromagnetic,barker2016static,bessarab2019stability,velkov2016phenomenology,jin2016dynamics,gobel2017antiferromagnetic,akosa2018theory}, 
FM skyrmions coupled antiferromagnetically through a spacer, so-called synthetic AFM skyrmions, were realized in multilayers~\cite{dohi2019formation,legrand2020room,finco2021imaging,juge2022skyrmions,chen2022controllable} while more complex AFM topological objects were identified in bulk materials~\cite{chmiel2018observation,gao2020fractional,Jani2021}. Ab-initio simulations predicted the emergence of intrinsic AFM skyrmions in Cr films and frustrated multimeronic states in Mn films interfaced with Ir(111) surface~\cite{aldarawsheh2022emergence,aldarawsheh2023intrinsic}. Intrinsic means that the AFM solitons are hosted within the same AFM material.

\begin{figure}[h!]
\centering
\includegraphics[width=1\linewidth]{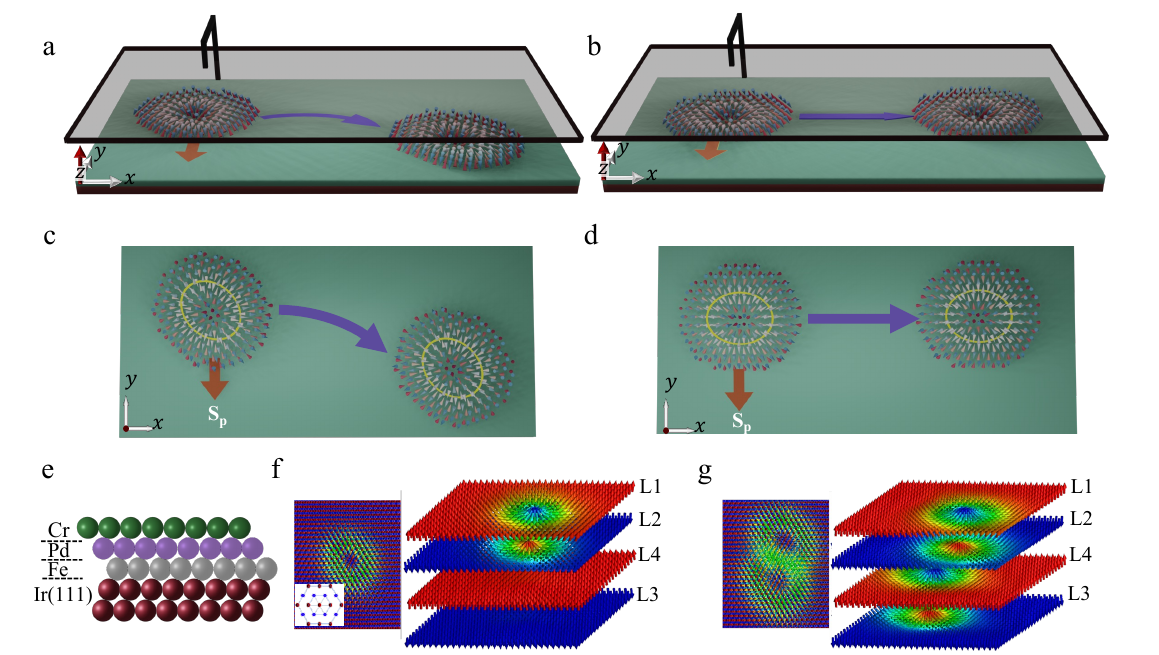}
\caption{\textbf{Current-driven dynamics of AFM skyrmions.} \textbf{a, b} Schematic representations of CPP induced motion of an elliptical AFM skyrmion showing the SkHE \textbf{a} or not \textbf{b} depending on the alignment of the polarization of the applied current $\mathbf{S}_P$ with respect to the skyrmion. \textbf{c, d} Top view of \textbf{a} and \textbf{b}, respectively. \textbf{e} Schematic representation of the material hosting intrinsic AFM skyrmions at the triangular lattice of a Cr layer grown on PdFe film deposited on an fcc(111) surface of Ir. The ground state in the Cr layer is the row-wise AFM configuration illustrated in the top view of the surface as red and blue spheres for different orientation of the spins shown in inset of \textbf{f}.  Snapshots of single \textbf{f} and double \textbf{g} overlapping AFM skyrmions  emerging in the Cr film with the spins distribution among four sublattices L1-L4. }
\label{fig:7_1}
\end{figure}

Here, we explore the dynamical response of intrinsic AFM skyrmions to an applied current. We consider the scenario of a magnetic tunnel junction (MTJ), where a magnetic electrode injects a perpendicular-to-plane  spin-polarized current (SP-CPP) with in-plane polarization on CrPdFe thin films deposited on Ir(111) surface  predicted from first-principles to host intrinsic AFM skyrmions~\cite{aldarawsheh2022emergence}(Fig.~\ref{fig:7_1}). Counter-intuitively, we demonstrate that these skyrmions exhibit a significant SkHE, which is strongly anisotropic, i.e. that is dependent on the polarization direction of the applied spin-current. We identify the origin of the SkHE and its vanishing conditions, which permits the design of track with or without the SkHE. We unveil complex interactions when interfacing intrinsic AFM skyrmions hosted in Cr with the  spin-textures, including individual FM skyrmions, found in Fe. This unique hybrid scenario enables the exploration of AFM-FM inter-skyrmion dynamics. The mutual inter-skyrmion interactions design a non-trivial two-dimensional energetical map, with pinning and repulsive centers, which impact on both the trajectory and velocity of AFM skyrmions and provide pinning and deflection processes. These findings pave the way for further exploration and control of skyrmion-based devices and applications in antiferromagnetic storage systems.

\section{Results}

\subsection{Trajectories of AFM skyrmions driven by perpendicular-to-plane currents.}

We investigate current-induced dynamics of  single and interchained   AFM skyrmions (Fig.~\ref{fig:7_1} f-g),  emerging at the Cr layer when deposited on   PdFe/Ir(111) with fcc stacking (Fig.~\ref{fig:7_1} e). PdFe/Ir(111) is a well established thin film material that hosts spin spirals, which can be deformed into FM skyrmions upon application of a magnetic field ~\cite{romming2013writing,dupe2014tailoring,Simon2014,romming2015field,crum2015perpendicular,fernandes2018universality,fernandes2020defect,Bouhassoune2021,lima2022spin}. Once covered with a Cr layer, a row-wise AFM (RW-AFM) state emerges, which hosts individual or catenated AFM skyrmions~\cite{aldarawsheh2022emergence} even without applying an external magnetic field. RW-AFM states in a triangular lattice are not common in nature~\cite{spethmann2020discovery,spethmann2021discovery}, and arise due to the interplay of  the neighboring Heisenberg exchange interactions ($J$)~\cite{kurz2001non,Aldarawsheh2023}. The presence of both the DMI and an out-of-plane magnetic anisotropy (K) stabilizes the highly sought intrinsic AFM skyrmions.  With the magnetic interactions among Cr atoms calculated from ab-initio (see Method section) and depicted in Supplementary Figure 1, the spins in the Cr layer are distributed among four sublattices (L1-L4). Ferromagnetically aligned spins on one sublattice host a single FM skyrmion, which is coupled to FM skyrmions emerging in  other sublattices to form AFM skyrmions. Single and interchained AFM skyrmions can form by populating in different fashion the distinct sublattices (Fig.\ref{fig:7_1} f-g).  Here we employ atomistic spin dynamics simulations, applying  Landau-Lifshitz-Gilbert (LLG) equation augmented with spin transfer torque (STT) terms~\cite{SLONCZEWSKI1996L1,chureemart2011dynamics,zhang2016magnetic,zhang2016antiferromagnetic,schurhoff2016atomistic}, as described in the Methods section, to explore the current-driven motion of the AFM skyrmions, by injecting perpendicular to plane currents with in-plane polarization (Fig.~\ref{fig:7_1} a-d). In this scenario, one expects a straight motion of an AFM skyrmion along the direction perpendicular to the polarization $\mathbf{S}_p$ of the applied spin current $\mathbf{j}_s$~\cite{ zhang2016magnetic,zhang2016antiferromagnetic}, as illustrated in Fig.\ref{fig:7_1} b-d.

\begin{figure}[h!]
\centering
\includegraphics[width=1.1\linewidth]{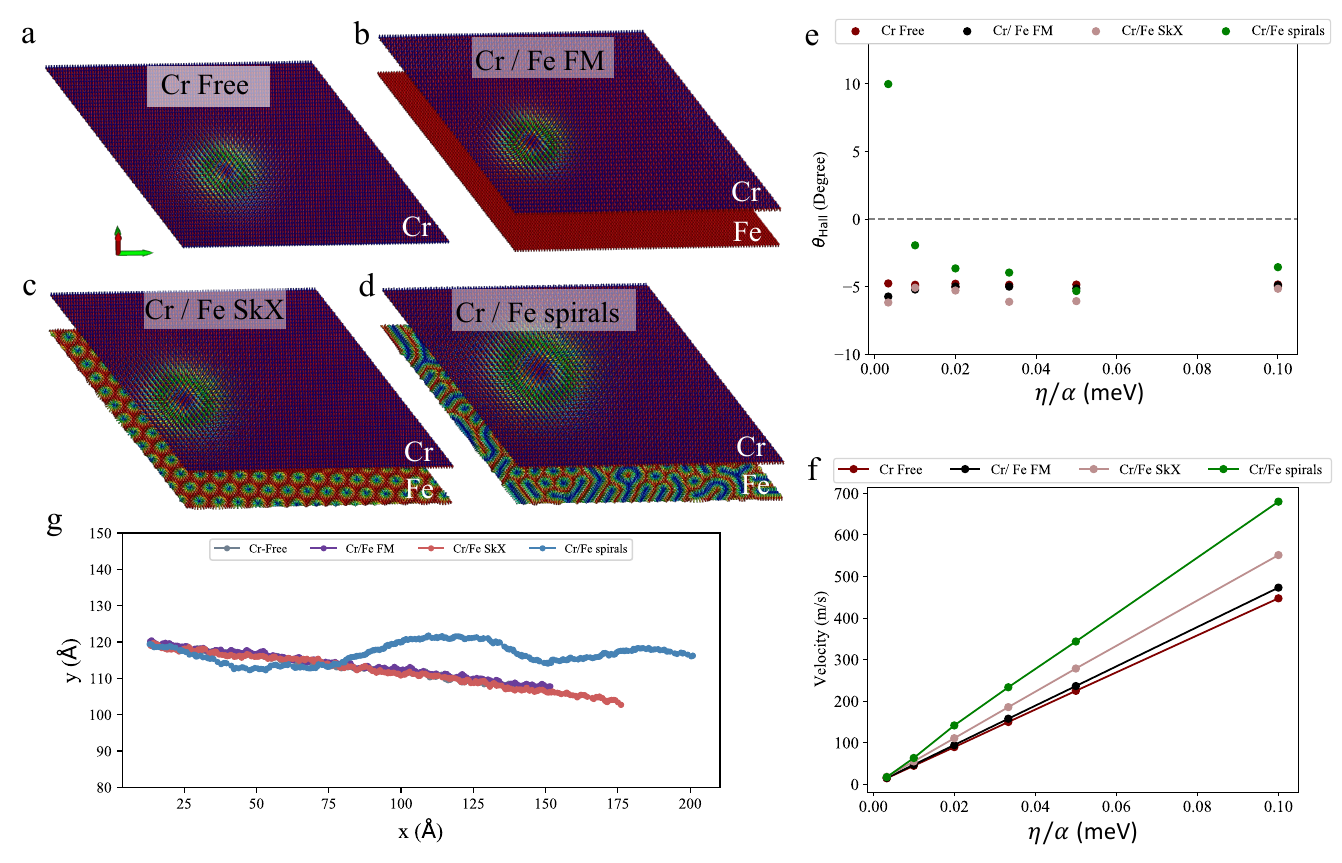}
\caption{\textbf{Transitional motion of elliptical AFM skyrmion driven by  SP-CPP currents.}  \textbf{a-d} Snapshots of the AFM skyrmion at Cr layer in four different cases: \textbf{a} Cr-free case, where the magnetic interactions with Fe layer are not included (AFM skyrmion radius is 2 nm); \textbf{b} Fe interactions are included with a finite magnetic field saturating Fe into a FM state (AFM skyrmion radius is 2.1 nm)  while \textbf{c} a weaker magnetic field leads to a skyrmion lattice (SkX) which slightly enlarges the AFM skyrmion (radius of 2.5 nm);  \textbf{d} in the absence of a magnetic field spirals emerge at Fe layer (AFM skyrmion reaches a  radius of 3.2 nm). \textbf{e} Impact of the current parameter ratio $\eta/\alpha$ on the skyrmion Hall angle, and \textbf{f} on velocity. \textbf{g} The trajectories of the AFM skyrmion for cases in \textbf{a-d} with $\eta/\alpha$ =0.01 meV.}
\label{fig:7_2}
\end{figure}

We initiate our study by investigating the case of single AFM skyrmions (Fig.~\ref{fig:7_1} a-d) and consider two possibilities:  either (i) by neglecting the Cr-Fe magnetic exchange interactions, which corresponds to a free standing Cr film (Fig.~\ref{fig:7_2} a), or (ii) not by scrutinizing various magnetic states in PdFe, which can be tuned by applying a magnetic field (Fig.~\ref{fig:7_2} b-d). For the latter, we consider the case of a saturated FM state in the Fe film (Fig.~\ref{fig:7_2} b), which is obtained upon application of a large magnetic field, while a moderate field can transition the spin-spiraling state shown in Fig.~\ref{fig:7_2} d to a skyrmion lattice (SkX) illustrated in Fig.~\ref{fig:7_2} c~\cite{romming2013writing,romming2015field}. The effective impact of the spin current can be monitored via the current parameter  $\eta$, which is directly proportional to ${j}_{s}$, divided by the Gilbert damping $\alpha$ (see Method section).

 As aforementioned, AFM skyrmions exposed to spin-polarized currents via STT, typically do not experience a SkHE~\cite{zhang2016magnetic,zhang2016antiferromagnetic,xia2019current,barker2016static,dohi2019formation} while their velocity is expected to be proportional to $\eta/\alpha$ in the case of CPP injection~\cite{xia2019current}. Surprisingly, our AFM skyrmions exhibit an unexpected dynamical behavior, deviating from conventional expectations. Independently from the Cr-Fe interaction and the nature of the magnetic state pertaining to the PdFe film, the Hall angle is found overall to be around -5$^\circ$ (negative sign means the deviation is clockwise), which  remains consistent across various $\eta/\alpha$ values (Fig.~\ref{fig:7_2} e). Deviations occur, however, for weak driving forces (small $\eta/\alpha$) when Cr is  placed atop Fe spin spirals. Indeed, the AFM skyrmions display then an irregular 'Brownian-like' motion as depicted in Fig.~\ref{fig:7_2} g, due to uncontrolled scattering at various spin-textures emerging in Fe. In this particular case, the extraction of the Hall angle is not trivial, since the skyrmion trajectories are not straight. Impressively, the Fe spirals can strongly deflect the AFM skyrmions, which can lead to effective Hall angles that are larger than 10$^\circ$, as calculated up to average distances of about 90 nm. Overall, the skyrmion Hall angles are found to be the smallest (largest) atop the SkX Fe (Fe spirals).

 The velocity of the skyrmions is linear with $\eta/\alpha$, with the largest speed found when Fe host a SkX state (Fig.~\ref{fig:7_2}f). Intriguingly, the Cr-Fe interaction  in general favors large skyrmion velocities, which can be traced back to the size of the skyrmions. Indeed, the Cr-Fe interaction enlarges the diameter of the AFM skyrmion, which is known to increase its velocity~\cite{jin2016dynamics,lee2018spin}, as unveiled in the upcoming analysis. Examples of skyrmion trajectories are shown in supplementary Movies 1-3.

 Fig.~\ref{fig:7_2} e-f is just the tip of the iceberg. By scrutinizing the skyrmion dynamics as function of the direction of the applied current, we unveil a rich anisotropic responses:  both the skyrmion Hall angles and velocities are modified and we identify directions along which the SkHE cancels out.  Before discussing the anisotropic current-driven dynamical response, we briefly address the origin of this behavior, which is induced by the ellipticity of the  AFM skyrmions emerging in Cr/PdFe/Ir(111) surface. By carefully scrutinizing the AFM skyrmions, one can identify an  elliptical shape. For instance, the single AFM skyrmion shown in Fig.~\ref{fig:7_9}c has a major and minor axis of 2.2, and 1.8 nm, respectively. Upon formation of a double AFM skyrmion, the shape of the skyrmions remains elliptical. The size of the skyrmions forming the solitonic dimer, experiences a significant increase, enlarging both the major and minor axes of the skyrmion building-blocks to 3 nm and 2.4 nm, respectively. The origin of the observed ellipticity can be traced back to the anisotropic magnetic environment experienced by the spins residing in the AFM skyrmion due to the interplay between the neighboring exchange interactions and the triangular lattice, which can be decomposed into four sublattices carrying distinct magnetic states  (a more detailed analysis is provided in Supplementary Note 1 and Supplementary Figure 2). Phenomenologically, one can demonstrate that by tuning the underlying interactions, the skyrmions can be reshaped into an isotropic form~\cite{Aldarawsheh2023}. We note that this is clearly a material dependent property.

\subsection{Directionality of the current-driven elliptical AFM skyrmions.}
\label{double_Hall}

\begin{figure}
\centering
\includegraphics[width=1\linewidth]{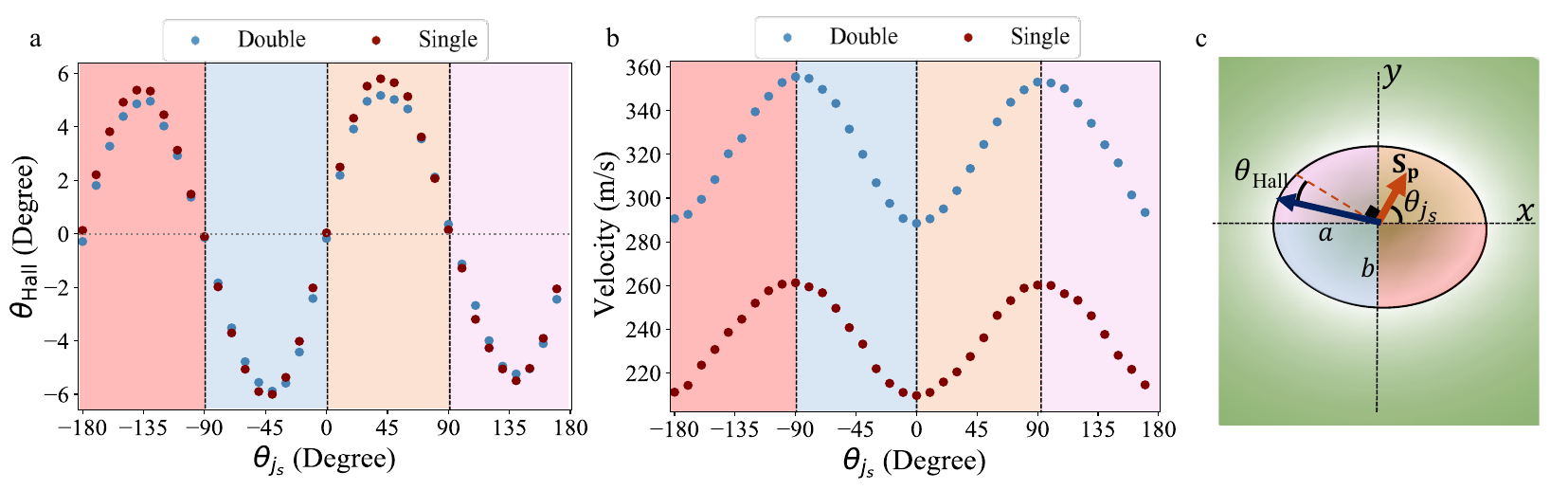}
\caption{\textbf{Influence of the direction of current polarization on the skyrmion Hall angle and velocity of elliptical AFM skyrmions.} \textbf{a} The skyrmion Hall angle and \textbf{b} velocity of AFM skyrmions as a function of  the angle $\theta_{j_s}$ between the current polarization direction and the major ellipse axis, for single (brown) and double(blue) AFM skyrmions, with $\eta/\alpha=$ 0.05 meV. \textbf{c} Schematic representation of the AFM skyrmion showing an elliptical shape, with a long (short) axis defining the $x$-axis ($y$-) axis. The angles associated with the spin-polarization of the current $\mathbf{S}_\mathrm{P}$ and the SkHE are displayed. The colored regions within the ellipse correspond to those shown in \textbf{a} and \textbf{b}.}
\label{fig:7_9}
\end{figure}

To explore the anisotropic current-driven response of the AFM skyrmions, we focus here on the case of the free-standing Cr layer, i.e. with the Cr-Fe interaction switched-off, which is also representative of the behavior found when the interaction is switched on while the Fe film hosts either the skyrmion lattice or the saturated FM state. As an example, we inject a current with $\eta/\alpha= 0.05$ meV, but varying systematically the angle $\theta_{j_s}$  between the in-plane current and the major axis of the ellipse, which is represented by the orange dashed line in Fig.~\ref{fig:7_9}c. 

The skyrmion Hall angle as function of $\theta_{j_s}$ is illustrated in Fig.~\ref{fig:7_9}a, which clearly shows an oscillating behavior for both the single (brown) and double (blue)  AFM skyrmions,  with color coded regions corresponding  to colored areas depicted in Fig.~\ref{fig:7_9}c. From Fig.~\ref{fig:7_9}a, one can notice that $\theta_{\text{Hall}}$ is suppressed when the current is polarized  along the two ellipse axes, and reaches its maximum value of about 6$^\circ$  when $\theta_{j_s}$ = 42$^\circ$.  Notably, it is not only the Hall angle that changes with $\theta_{j_s}$; the velocity of  AFM skyrmions also varies as shown in Fig.~\ref{fig:7_9}b, exhibiting the maximum (minimum) velocity when the skyrmions move along the ellipse major (minor) axis.  Interestingly as depicted in Fig.~\ref{fig:7_9}a, double and single AFM skyrmions  show the same Hall angle when subjected to the same polarized currents, however the double AFM skyrmion moves faster than the single (Fig.~\ref{fig:7_9}b).

\subsection{Thiele equation for elliptical AFM skyrmions.}
To comprehend these intriguing findings, we analyze the Thiele equation governing AFM skyrmions driven by CPP~\cite{Tomasello2013, zhang2016magnetic,zhang2016antiferromagnetic, jiang2017direct} (see more details in Supplementary Note 2):
\begin{equation}
\label{equ.thiele-CPP}
-\alpha \mathcal{D} \cdot \textbf{v}+ \mathcal{B} \cdot \textbf{S}_p=0,
\end{equation} 
where $\textbf{v}$ denotes the skyrmion velocity while  $\textbf{S}_p = (\cos\theta_{j_s},\sin \theta_{j_s})$ stands for the unit vector defining the spin polarization direction of the injected current. Note that the $x$ and $y$ axes are defined respectively by the semi-major and semi-minor axes of the elliptical skyrmion (see Fig.~\ref{fig:7_9}c). $\mathcal{B}$ is the driving force related tensor, and $\mathcal{D}$ is the dissipative tensor for the whole AFM skyrmion. If the skyrmions were isotropic in shape, the velocity would be perpendicular to the polarization of the current and there would not skyrmion Hall effect.

 For each building-block FM skyrmion $i$  forming the AFM skyrmion, the associated components of the dissipative tensor~\cite{xia2020dynamics}, assuming skyrmions of identical size and shape, are given by: 
 $(\mathcal{D}^i_{xx},\mathcal{D}^i_{yy}) = \frac{\pi^{2}}{8} \left(\frac{b}{a},\frac{a}{b}\right) 
 $,  where $a$, and $b$ are the semi-major and semi-minor ellipse axes, while $\mathcal{D}^i_{yx}=\mathcal{D}^i_{xy}=0$. The  components of the driving force tensor are  $(\mathcal{B}^i_{x y},\mathcal{B}^i_{y x})= \frac{\gamma \pi}{8\mu_s} \eta (- b,a)$ 
  and $\mathcal{B}^i_{x x}=\mathcal{B}^i_{y y}=0$. The skyrmion velocity is then given by:
  \begin{eqnarray}
  \label{eq.velocity}
    \textbf{v} &=& \frac{1}{\alpha}\left(\frac{\mathcal{B}_{xy} }{ \mathcal{D}_{x x}}\sin \theta_{j_s},\frac{\mathcal{B}_{yx} }{\mathcal{D}_{y y}}\cos \theta_{j_s}\right) =\frac{\gamma \eta a}{\pi \mu_s \alpha}\left(-\sin \theta_{j_s},\frac{b}{a}\cos \theta_{j_s}\right) ,\\
    |\textbf{v}| &=& \frac{\gamma \eta a}{\pi \mu_s \alpha} \sqrt{\sin^2{\theta_{j_s}} + \frac{b^2_{sk}}{a^2_{sk}}\cos^2{\theta_{j_s}} } ,
  \end{eqnarray}
where one immediately notices that if the skyrmions were circular isotropic, the polarization of the spin-current is perpendicular to the velocity since $\textbf{v}\cdot\textbf{S}_p = 0$. The propagation direction associated to the isotropic case case defines the reference angle from which the skyrmion Hall angle is measured $\theta_\text{ref} = \arctan{\frac{v_{y}}{v_{x}}} = \theta_{j_s} + \frac{\pi}{2}$. Therefore, the Hall angle is evaluated from
\begin{eqnarray}
    \theta_{\text {Hall}} &=&\arctan{\left[ \frac{b}{a} \tan \left( \theta_{j_s}+\frac{\pi}{2}\right)\right]} - \left(\theta_{j_s} + \frac{\pi}{2}\right)\,.
\end{eqnarray}

With these findings at hand, we can explain the behavior of the AFM skyrmions. If the current is polarized along the major or minor axes of the spins, i.e. $\theta_{j_s}$ is a multiple of $\frac{\pi}{2}$, the Hall angle cancels out, which define the extrema of the velocity given by $\frac{\gamma \eta b}{\pi \mu_s \alpha}$ for $\theta_{j_s} = 0, \pi$ (minimum) and $\frac{\gamma \eta a}{\pi \mu_s \alpha}$ for $\theta_{j_s} = \frac{\pi}{2}, \frac{3\pi}{2}$ (maximum). The ratio $\frac{b}{a}$, which is about 0.8 for both the single and double AFM skyrmions, shapes the magnitude of the Hall angle as well as the range of oscillations in the velocity. 
 This means with elliptical AFM skyrmions, we have two more degrees of freedom to manipulate the CPP induced motion of the AFM skyrmions, where the skyrmion exhibits its maximum velocity when injecting currents polarized along its minor axis, resulting in a Hall free motion along the major axis. This analysis goes along with our findings depicted in Fig.~\ref{fig:7_9} b, where the double AFM skyrmion with  dimensions of $(a,b) = (3, 2.4)$\,\text{nm} moves with maximum velocity of 355  while the single AFM skyrmion with smaller size (dimensions of $(2.2, 1.8)$ nm), and hence slower motion according to eq.~\ref{eq.velocity}, where its  maximum velocity reaches 260 m/s. 
The maximum Hall angle is expected for $\theta_{j_s} + \pi/2 = \arccos{\sqrt{\frac{b}{a+b}}}$, which leads to $\theta_{j_s} = 42^\circ$  and $\theta_\text{Hall} = 6.2^\circ$ in agreement with the numerical findings of the previous section.  
 Notably, the impact of ellipticity on the motion of FM skyrmions subjected to spin currents was discussed in Refs.~\cite{huang2017stabilization,xia2020dynamics,chen2018dynamics}.

\subsection{Current-driven dynamics of AFM skyrmions interacting with FM skyrmions.}
\begin{figure}[h!]
\centering
\includegraphics[width=1.0\linewidth]{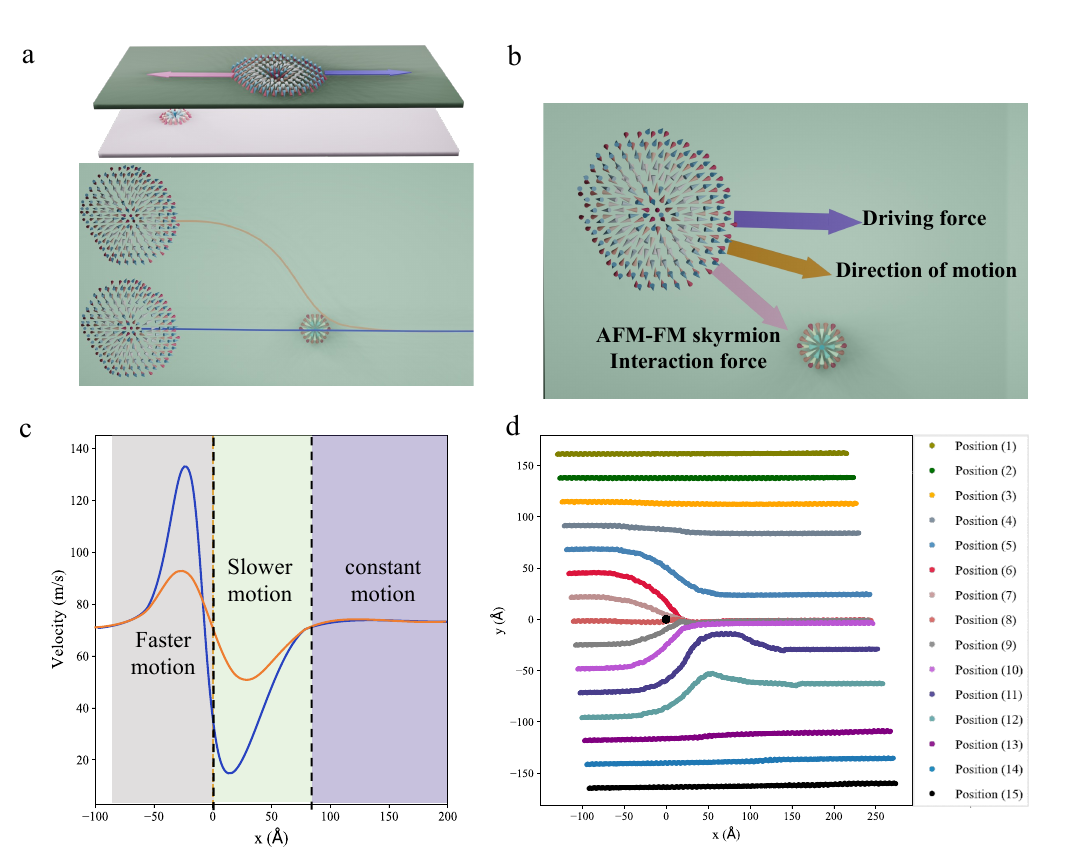}
\caption{\textbf{Position dependent deflection of AFM skyrmions due to the  AFM-FM skyrmionic interaction. } 
\textbf{a, b} Schematic representation of the forces acting on the AFM skyrmion:  The driving force due to the applied spin current and the AFM-FM skyrmionic interacting induced by the FM skyrmion in the Fe layer through the Pd-spacer. In \textbf{a} two trajectories with different starting points are illustrated. \textbf{c} Schematic representation of the effect of the Fe FM skyrmion on the velocity of the AFM skyrmion with the blue and orange lines corresponding to the paths shown in \textbf{a}.  \textbf{d} The trajectory of the  AFM skyrmion shown in \textbf{a} when current-driven toward a pinned FM skyrmion considering different initial positions (motion from left to right). Deflection in the motion direction occurs depending on the relative position between the AFM and FM skyrmions.} 
\label{fig:7_6}
\end{figure}

When the Fe substrate hosts spirals with an inhomogeneous  distribution of FM skyrmions, we unveiled in Fig.~\ref{fig:7_2}g  that the intrinsic AFM skyrmions driven in the Cr overlayer exhibit typical dynamics pertaining to interactions with defects. In this section, we explore the synthetic configuration of an AFM skyrmion interacting with a FM one through a Pd film. This scenario is trivially realized in Cr/PdFe/Ir(111) surface by applying an out-of-plane magnetic field, which reduces the size of the FM skyrmions in Fe and transforms the lattice configuration into individual topological objects. By applying spin-polarized current, as done previously, we drive an AFM skyrmion living in the Cr film towards a pinned FM skyrmion hosted by the Fe layer. We consider two cases, either the planned skyrmion trajectory passes trough the FM skyrmion, or it is shifted (see Fig.~\ref{fig:7_6} a-b). 

We notice that the AFM skyrmion gets pinned at the FM skyrmion for a weak applied current, see example $\eta/\alpha =$ 0.001 meV in Supplementary movie 4, which clearly indicates the attractive nature of the FM-AFM skyrmion interaction. A stronger current, e.g. $\eta/\alpha$ = 0.017 meV, enables the AFM skyrmion to escape the trapping FM soliton following a roller-coaster-like motion, with a velocity increase from the initial 75 m/s to reach  130 m/s  once getting close to the FM skyrmion, which leads to a "speeding up zone"  as shown in Fig.~\ref{fig:7_6}c.  Once the AFM skyrmion overtakes the FM skyrmion, the velocity reduces by about  88$\%$ down to around 15 m/s due to the FM-AFM skyrmion interaction that opposes the driving force and leads to a "slowing-down zone" . As the AFM skyrmion moves away from the FM skyrmion, only the driving force dictates its motion, resulting in a "constant motion regime," depicted in Fig.~\ref{fig:7_6} c and Supplementary Figure 3, where the velocity stabilizes at approximately 75 m/s.  

If the AFM skyrmion is off-centered with respect to the FM one, their mutual attractive interaction is capable of deviating the underlying trajectory to bring the AFM skyrmion to the vicinity of the FM one  as depicted schematically in Fig.~\ref{fig:7_6}b and demonstrated systematically for different paths illustrated in Fig.~\ref{fig:7_6}d  and Supplementary Figures 3-5. Due to the AFM-FM skyrmion attraction the AFM skyrmion deflects at the vicinity of the FM one and then continues its motion along a straight line with a velocity of 75 m/s. Intriguingly, a second deflection manifests when the AFM skyrmion starts an approach from positions (11) and (12) after passing the FM skyrmion, which signals a non-trivial energy profile of the hybrid AFM-FM skyrmionic interaction.

\subsection{AFM-FM skyrmion interaction profile.}

\begin{figure}[h!]
\centering
\includegraphics[width=1.0\linewidth]{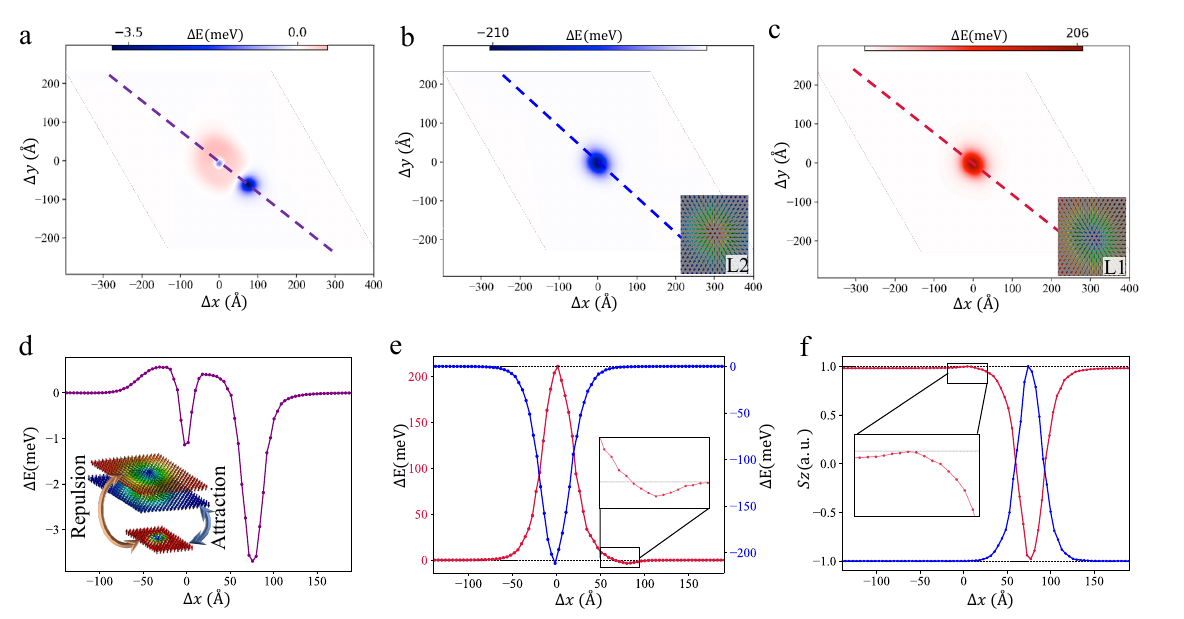}
\caption{\textbf{AFM-FM skyrmion interaction profile.} 
\textbf{a} Two-dimensional heat map of the total energy difference resulting when rigid-shifting the AFM skyrmion all over the lattice with the presence of Fe-FM skyrmion at the center of the Fe layer. The energy difference is taken with respect to the case when the two skyrmions are not interacting. This energy difference is further decomposed into the interaction with the AFM skyrmions building-blocks: the red-cored FM skyrmion residing at sublattice L2 \textbf{b}, which is of attractive nature, and the interaction with the blue-cored skyrmion residing at sublattice L1 \textbf{c}, which is of repulsive nature.  \textbf{d} Energy profile along the purple line indicated in \textbf{a}, decomposed in \textbf{e} into the L2 blue and L1 red  contributions. \textbf{f} The z component of the spin in the two sublattices L1 (blue) and L2 (red) along the purple line when the AFM skyrmion positioned at the second minimum shown in \textbf{a}. Inset in \textbf{d} is a schematic representation of the nature of the interaction  between the building blocks of the AFM skyrmion and the  FM Fe-skyrmion.  }
\label{fig:7_66}
\end{figure}

To elucidate the underlying reason for the unanticipated second deflection mentioned in the previous section, we analyse   the AFM-FM skyrmion interaction profile. Since the AFM skyrmion in Cr film is made of two FM skyrmions oppositely oriented with respect to each other, we expect two competing interactions with the FM skyrmion in Fe layer, as shown in the inset of  Fig.~\ref{fig:7_66} d. The FM skyrmions having their cores pointing in the same direction and residing in the same ferromagnetic background would repel each other, while those having an opposite magnetic alignment would attract each other.  This is clearly illustrated in Fig.~\ref{fig:7_66} b, and c, respectively, which shows the energy contribution of both types of coupling to the energy profile, that is obtained by rigidly shifting the AFM skyrmion all over the Cr lattice atop the FM skyrmion pinned in center of the Fe film. The total interaction heat map depicted in Fig.~\ref{fig:7_66} a exhibits a minimum when the AFM skyrmion overlaps with the FM skyrmion, signifying their mutual attraction. A second minimum appears when the AFM skyrmion is positioned in the lower right part of the FM skyrmion, which explains the aforementioned second deflection experienced by the AFM skyrmion as noted in the previous section.

The surprising second minimum finds its origin in the intrinsic asymmetric shape of the AFM skyrmion with respect to the skyrmion core residing in one of the sublattices, as illustrated in Supplementary Figure 6. 
When interfaced with the Fe FM spins, the Cr spins residing in the background of  sublattice L1 tilt away from their initial direction, while those residing in sublattice L2 get more collinear and antiferromagnetically aligned with respect to the Fe-magnetization, see the red and blue plots respectively in Fig.~\ref{fig:7_66} f obtained along the purple line in Fig.~\ref{fig:7_66} a when positioning the Cr skyrmion at the second minimum. At the vicinity of the Fe FM-skyrmion, the interaction picture gets reversed, which leads to the sublattice-dependent interaction profile shown in Fig.~\ref{fig:7_66} e. In particular, the asymmetric profile of the AFM skyrmions together with the magnetic interaction across the sublattices enable an energy gain in an area of sublattice L1, where the Cr spins benefit from the AFM coupling with the core of the Fe skyrmion, see the kink in the spin profile highlighted in the inset of Fig.~\ref{fig:7_66} f. Unveiling the interaction profile between AFM and FM skyrmions holds significant importance, as it offers an opportunity for manipulating and regulating the trajectories and dynamics of  AFM skyrmions by strategically positioning pinned FM skyrmions at the Fe layer.

\section{Discussion}
In this study, we uncovered  the intricate dynamics of intrinsic AFM skyrmions subjected to perpendicular to plane spin polarised currents, with a particular attention to the impact of FM skyrmions emerging in a hybrid heterostructure (Cr/PdFe/Ir(111) surface) made of  an AFM layer (Cr) separated from a FM layer (Fe) by a Pd spacer layer. 
In contrast to expectations, even in AFM skyrmions we demonstrate the emergence of the SkHE stemming from the elliptical shape of the topological states. Both the SkHE and skyrmion velocity are anisotropic and follow well defined dependencies with respect to the polarization direction of the applied currents. The ability to manipulate the polarization direction of impinging spin currents provides a clear avenue for designing tracks where the skyrmion Hall effect (SkHE) either diminishes or persists.  
The presence of non-trivial magnetic states in the FM film can impact the dynamics of the AFM skyrmions by tuning both their velocity and trajectory. For instance, FM skyrmions act as pinning centers  , which depending on the applied current can deflect AFM skyrmions. The seeding of FM skyrmions modifies non-trivially the emergent hybrid AFM-FM skyrmionic interaction profile, which can host several minima, offering the potential of customizing pathways for the motion of AFM skyrmions (see examples illustrated in Supplementary Figure 7).

In summary, our study advances the understanding of AFM skyrmion dynamics and their interplay with FM skyrmions. These insights hold great promise for the development of innovative spintronic devices that harness the unique properties of AFM and FM spin textures. As the field of AFM spintronics continues to evolve, this research contributes to the foundation for efficient information processing and storage schemes, potentially revolutionizing the realm of next-generation spintronic technologies.

\begin{methods}
\label{ch:dynamics_computational}
To simulate the trilayer of CrPdFe deposited on Ir(111) with fcc stacking, we employ DFT. To relax the magnetic layers we use the Quantum-Espresso computational package\cite{giannozzi2009quantum}. The projector augmented wave pseudo potentials from the PS Library \cite{dal2014pseudopotentials} and a $28\times28\times1$  k-point grid were used for the calculations. Then the electronic structure and magnetic properties were simulated using all electron full potential  relativistic Koringa -Kohn-Rostoker (KKR) Green function method \cite{Papanikolaou2002,Bauer2014,russmann2022judftteam} in the local spin density approximation. The Heisenberg exchange interactions and DM vectors were extracted using the infinitesimal rotation method~\cite{Liechtenstein1987,Ebert2009} with a k-mesh of a $200\times200$. More details on the simulation procedure can be found in Ref.~\cite{aldarawsheh2022emergence}.

\subsection{Spin atomistic dynamics:}

To explore the intricate dynamics of the AFM skyrmions, we conducted spin atomistic simulations. In our study, we consider a two dimensional  Heisenberg model on a triangular lattice, equipped with Heisenberg exchange coupling, DMI, the magnetic anisotropy energy, and Zeeman term. All  parameters were obtained from ab-initio. The energy functional reads as follows:

 \begin{equation}
   \mathcal{H} = \mathcal{H}_\text{Exc } + \mathcal{H}_\text{ DMI} + \mathcal{H}_\text{Ani} + \mathcal{H}_\text{Zeem},
   \label{eq.Heisenberg}
\end{equation}
with:
\begin{equation*}
 \mathcal{H}_\text{Exc}= -\sum\limits_{<i,j>} J^\text{Cr-Cr}_{ij}\; \textbf{S}_{i}\cdot \textbf{S}_{j}       -\sum\limits_{<i,j>} J^\text{Fe-Cr}_{ij}\; \textbf{S}_{i}\cdot \textbf{S}_{j}   -\sum\limits_{<i,j>} J^\text{Fe-Fe}_{ij}\;\textbf{S}_{i}\cdot \textbf{S}_{j},
\end{equation*}
\begin{equation*}
 \mathcal{H}_\text{DMI}= \sum\limits_{<i,j>}\textbf{D}^\text{Cr-Cr}_{ij}\cdot [\textbf{S}_{i}\times \textbf{S}_{j}]+\sum\limits_{<i,j>}\textbf{D}^\text{Fe-Cr}_{ij}\cdot [\textbf{S}_{i}\times \textbf{S}_{j}]+\sum\limits_{<i,j>}\textbf{D}^\text{Fe-Fe}_{ij}\cdot [\textbf{S}_{i}\times \textbf{S}_{j}],  
 \end{equation*}
 \begin{equation*}
 \mathcal{H}_\text{Ani}=- K^\text{Cr}\sum\limits_{i}  (S_i ^z)^2 - K^\text{Fe}\sum\limits_{i}  (S_i ^z)^2,
 \end{equation*}
 \begin{equation*}
\mathcal{H}_\text{Zeem} =- \sum\limits_{i} h_i S_i^z,  
\end{equation*}
  where $i$ and $j$ are site indices carrying each magnetic moments.  $\textbf{S}$ is a unit vector of the magnetic moment. $J^\text{X-Y}_{ij}$ is the Heisenberg exchange coupling strength, being $<$ 0 for AFM interaction, between an X atom on site $i$ and a Y atom on site $j$. A similar notation is adopted for the DMI vector $\textbf{D}$ and the magnetic anisotropy energy $K$. The latter favors the out-of-plane orientation of the magnetization. In this study, in order to promote the formation of a single AFM skyrmion, independent of the Fe magnetic state's influence; we reduced the Cr atom magnetic anisotropy  from 0.5 to 0.4 meV per magnetic atom.  $h_i= \mu_i B$  describes the Zeeman coupling to the atomic spin moment $\mu$ at site $i$ assuming an out-of-plane field.

\subsection{Landau–Lifshitz–Gilbert (LLG) equation for CPP.}
 We  apply the spin-polarized CPP injection to induce transitional motion of  AFM skyrmions. In the CPP case, the  current is
perpendicular to the film plane, but polarized in  in-plane direction. The dynamics of the magnetization $\mathbf{\textbf{S}}_i$ at the lattice site $i$ is then  governed by the extended LLG equation taking into account the STT term~\cite{SLONCZEWSKI1996L1,chureemart2011dynamics,zhang2016magnetic,zhang2016antiferromagnetic,schurhoff2016atomistic},

\begin{equation}
\begin{aligned}
\frac{\mathrm{d} \mathbf{\textbf{S}}_i}{\mathrm{~d} t}=&-\frac{\gamma}{(1+\alpha^2)\mu_i}\mathbf{\textbf{S}}_i \times \mathbf{\textbf{B}}_{\text {eff }}^i- \frac{\gamma\alpha}{(1+\alpha^2)\mu_i}\mathbf{\textbf{S}}_i \times\left(\mathbf{\textbf{S}}_i \times \mathbf{\textbf{B}}_{\text {eff }}^i\right) \\
&+ \frac{\gamma\alpha \eta}{(1+\alpha^{2})\mu_B}\mathbf{\textbf{S}}_i \times\mathbf{\textbf{S}}_p- \frac{\gamma \eta}{(1+\alpha^2)\mu_B}\mathbf{\textbf{S}}_i \times\left(\mathbf{\textbf{S}}_i \times \textbf{S}_p\right),
\end{aligned} 
\label{LLG_current}
\end{equation}

where $\gamma$ is the gyromagnetic ratio,  $\alpha$  is the Gilbert damping controlling the dissipation
of angular momentum and energy from the magnetic subsystem, $\mathbf{\textbf{B}}_{\text {eff }}^i$ is the effective field related to the energy gradient  $(-\frac{\partial \mathcal{H}}{\partial \mathbf{\textbf{S}}_i})$, and $\textbf{S}_p$ is the current polarisation direction with the current amplitude   monitored by the parameter $\eta=\frac{j_s Pg\mu_{B}}{2ed M_s\gamma}$,  with the absolute value of the current density $j_s$ , polarisation $P$, saturation magnetisation $M_s$, the Land\'e $g$-factor $g$,
magnitude of electron charge $e$ and Bohr magneton $\mu_{B}$, d is the film thickness, and $\gamma$ is the gyromagnetic ratio. \\
In our study, we utilize the LLG equation as implemented in the Spirit code~\cite{muller2019spirit}. We assumed periodic boundary conditions to model the extended two-dimensional system with cells containing  $200^2$. 
\end{methods}
\subsection{Data availability}
The data that support the findings of this study is  available from the corresponding author upon reasonable request.

 \subsection{Code availability} We used  SPIRIT code which can be found at \url{https://github.com/spirit-code/spirit}. Additionally, the KKR code, a sophisticated ab-initio DFT tool, can be found at \url{https://github.com/JuDFTteam/JuKKR}.

\begin{addendum}

\item  This work was supported by the Federal Ministry of Education and Research of Germany
in the framework of the Palestinian-German Science Bridge (BMBF grant number
01DH16027) and the Deutsche For\-schungs\-gemeinschaft (DFG) through SPP 2137 ``Skyrmionics'' (Projects LO 1659/8-1). 
The authors gratefully acknowledge
the computing time granted through JARA on the supercomputer JURECA %\cite{ krause2018jureca}  
at Forschungszentrum Jülich.  M.S. acknowledges the fund from the European Research Council (ERC) under the European Union’s Horizon 2020 research and innovation program (Grant No. 856538, project “3D MAGiC”.

\item[Author contributions] S.L. initiated, designed and supervised the project. A.A. performed the simulations and analysed the results with the supervision of S.L.. All authors discussed the results. A.A. and S.L. wrote the manuscript to which all co-authors contributed.

\item[Competing Interest ] The authors declare no competing interests.

\item[Correspondence] Correspondence and requests for materials should be addressed to A.A. (email: a.aldarawsheh@fz-juelich.de) or to S.L. (email: s.lounis@fz-juelich.de).

\end{addendum}

\section*{References}
\bibliographystyle{naturemag}
\bibliography{references}

\end{document}